# Simultaneous Measurement of Rock Permeability and Effective Porosity using Laser-Polarized Noble Gas NMR

R. Wang[*,+], R. W. Mair[*,+], M. S. Rosen[*], D. G. Cory[+], and R. L. Walsworth[*]

[*] Harvard-Smithsonian Center for Astrophysics, Cambridge, MA 02138, USA.

[+] Massachusetts Institute of Technology, Cambridge, MA, 02139, USA.

**Corresponding Author:**

Ross Mair

Harvard Smithsonian Center for Astrophysics,

60 Garden St, MS 59,

Cambridge, MA, 02138,

USA

Phone: 1-617-495 7218

Fax: 1-617-496 7690

Email: rmair@cfa.harvard.edu

# ABSTRACT

We report simultaneous measurements of the permeability and effective porosity of oil-reservoir rock cores using one-dimensional NMR imaging of the penetrating flow of laser-polarized xenon gas. The permeability result agrees well with industry standard techniques, whereas effective porosity is not easily determined by other methods. This novel NMR technique may have applications to the characterization of fluid flow in a wide variety of porous and granular media.

**PACS numbers:**  32.80.Bx    Optical Pumping

## I. INTRODUCTION

Porous media are ubiquitous in nature. Examples include granular materials, foams, ceramics, animal lungs and sinuses, and oil- or water-bearing "reservoir" rocks. Diagnosing the structure of these materials is relevant to a wide range of scientific and technological problems. For example, knowledge of the fluid transport properties of reservoir rocks is important for the monitoring of contaminant percolation and for oil extraction. Similarly, knowledge of the evolution of the porous structure of materials subjected to large thermal or mechanical stress may help characterize the dynamics of cracking and material failure. There is a continuing need for the development and application of new techniques that characterize complex systems such as fluid flow in porous media.

Two of the most important parameters used to characterize porous media are permeability and effective porosity [1]. Permeability is a measure of the ability of a porous material to transmit fluid, and is defined by Darcy's law as the proportionality constant relating the volume flow rate, $\bar{q}$, for an incompressible fluid of viscosity $\mu$, to the pressure gradient, $\nabla P$, driving the flow [2]:

$$\bar{q} = -\frac{kA}{\mu}\nabla P, \quad (1)$$

where $k$ is the permeability and $A$ is the total cross-sectional area of the porous material. (Note that Darcy's law is valid only for linear laminar flow where the Reynolds number (Re), based on average pore diameter, does not exceed 1.0 [1]. In the experiments reported here, Re ~ $10^{-5}$.) Permeability is determined by measuring the fluid pressure difference across a sample, and the resulting flow rate through it. There are various techniques that realize this basic scheme, including measurements of gas flow [3].

In one class of porous media of great practical interest, reservoir rocks, permeability can vary greatly. For example, in sandstones, where the pores are large and well connected, the permeability is large: $k$ ~ 1 D [4], where 1 D = one darcy = 0.978 $\mu$m$^2$ [1]. Impermeable rocks, such as siltstones, consist of fine or mixed-sized grains, and hence have smaller or fewer interconnected pores with $k$ ~ 1 mD [4].

The total or absolute porosity, $\phi$, is simply the fractional volume of all void space inside a porous material, whether or not the voids are interconnected and make a continuous channel through the sample. More useful, when considering fluid flow, is the effective porosity $\phi_e$: the volume fraction of pore spaces that are fully interconnected and contribute to fluid flow through the material, excluding dead-end or isolated pores that are not part of a flow path. Effective porosity also relates the average fluid velocity, or Darcy velocity $v_d = \bar{q}/A$, to the mean velocity of a tracer flowing through the pore space, $v_s$, according to the simple relation [5]:

$$\phi_e = \frac{A_f}{A} = \frac{v_d}{v_s} = \frac{\bar{q}/A}{v_s} = \frac{-k \nabla P/\mu}{v_s}, \qquad (2)$$

where $A_f$ is the effective cross-sectional area of the sample where flow occurs. Although effective porosity can be defined in a number of ways, absolute porosity is always larger than or equal to the effective porosity for a given sample [5]. An accurate measure of effective porosity is important for understanding fluid flow in porous media, and phenomena such as the diffusion, dispersion and deformation of the solid phase due to stress resulting from the applied pressure.

Absolute porosity and permeability are readily measurable with existing techniques [6], although the two parameters generally need to be measured separately with different methods, and many of the techniques (e.g., mercury intrusion porosimetry), are either invasive, toxic, or both [6,7]. Moreover, effective porosity, the more informative porosity parameter for fluid transport in porous media, cannot generally be measured directly with current standard techniques [5,8].

In this paper we demonstrate the simultaneous measurement of permeability and effective porosity in reservoir rocks using laser-polarized noble gas NMR imaging, a powerful, non-invasive probe of the spatial distribution and motion of fluid inside a porous sample. NMR of gas-phase samples has traditionally been hampered by low nuclear spin density, ~ 3 orders smaller than for solid or liquid samples, which results in a much lower signal to noise ratio for thermally spin-polarized samples at the same magnetic field strength. Here, we use the spin-exchange optical pumping method [9] to enhance the nuclear spin polarization of $^{129}$Xe gas by 3 - 4 orders of magnitude, producing a magnetization density which can be as high as water samples at magnetic fields of ~ 1 tesla. An additional benefit of using a laser-polarized gas, especially for tracer studies like those used here, is the ability to induce a step-change in the magnetization with a train of saturating RF and gradient pulses [10,11]. As the gas polarization has been produced external to the main applied magnetic field, such a saturation train essentially sets the xenon magnetization to zero – with the only replenishment being from polarized gas that flows into the sample after magnetization saturation. The thermal (Boltzmann) polarization that re-establishes itself after the saturation train is so small as to be negligible. The ability to manipulate the xenon magnetization in this way is a key component in the measurements we will describe below.

To determine permeability and effective porosity, we monitored the movement of $^{129}$Xe spins through each rock sample by measuring the one-dimensional NMR signal profile, yielding a one-dimensional image of the spatial distribution of spin magnetization that is dependent on the characteristics of the porous medium. We measured steady-state $^{129}$Xe NMR profiles with the polarized gas flowing through the sample; we also measured penetration profiles for different inflow times. These 1D images were analyzed in terms of a well-known application of Darcy's Law to porous media that are homogeneous on large length scales ($\geq$ 100 $\mu$m in the present case) [4,12].

## II. EXPERIMENTAL PROCEDURE

Xenon gas (26.4% abundance of $^{129}$Xe) was spin-polarized in a glass cell which contained a small amount of Rb metal and a total gas pressure of ~ 4 bar, with ~ 92% xenon and the remainder $N_2$. We heated the cell to 105°C to create an appropriate Rb vapor density and induced spin polarization in the Rb vapor via optical pumping on the Rb D1 line (~ 795 nm) using ~ 12 W of broad-spectrum (~ 2 nm) light provided by a fiber-coupled laser diode array [8,13]. In about 5 minutes Rb-Xe collisions boost the $^{129}$Xe spin polarization to ~ 1%, and a continuous output of polarized xenon gas was then provided to a porous sample in an NMR instrument at a controlled flow rate of 50 cm$^3$ per minute. Fig. 1 shows a schematic of the experimental apparatus.

We employed both high and low permeability rocks in our demonstration measurement. The high permeability rock was Fontainebleau sandstone, a simple, homogeneous rock type that is largely free of paramagnetic impurities and has a regular and fairly narrow distribution of pore sizes (~ 10 to 100 $\mu$m) [12]. We also studied a low permeability rock, Austin Chalk, a very fine grained, spatially homogeneous rock with high porosity but very small (< 10 $\mu$m) and poorly connected pores. Additional rocks are being studied as part of an ongoing project – the results from these samples will be presented elsewhere. All rock samples were cylindrically shaped, with a diameter of 1.9 cm and a length of 3.8 cm. We baked the samples under vacuum before use to ensure absorbed water in the pore space was removed. The rock being probed was held in a sample cell primarily built of machined Teflon, inside of which the $^{129}$Xe spin relaxation time is $T_1 \sim 2$ minutes, much larger than the values measured in the pores of rock samples ($^{129}$Xe $T_1 \sim$ 1 - 10 seconds). The rock cell was connected to the xenon polarization chamber via 1/8 inch ID Teflon tubing, and the entrance to the rock cell also contained a diffuser plate made of 5 mm thick Teflon, with 46 holes each of diameter 1.2 mm, to distribute flowing xenon gas evenly into the sample. The exit side of the sample was connected via similar Teflon tubing to a vacuum pump that induced gas flow through the rock sample. The gas flow rate was regulated by a mass flow controller, placed just before the vacuum pump, which provided steady flows ranging from 10 to 1000 cm$^3$/s. In continuous flow mode, the gas moved from the supply bottles, through the polarization chamber and then the rock sample, and finally through the mass flow controller and on to the vacuum pump.

The rock sample was positioned in a 4.7 T horizontal bore magnet, interfaced to a Bruker AMX2-based NMR console. We employed an Alderman-Grant-style RF coil [Nova Medical Inc., Wakefield, MA] for $^{129}$Xe observation at 55.4 MHz. All NMR imaging experiments were non-slice-selective one-dimensional profiles along the flow direction employing a hard-pulse spin echo sequence with echo time, $t_E = 2.1$ ms and an acquired field of view of 60 mm. Steady-state flow profiles were obtained by this method without pre-saturation. The $^{129}$Xe polarization penetration depth was measured by preceding the echo sequence with a saturation train of RF and gradient pulses to destroy all $^{129}$Xe magnetization inside the rock sample; and then waiting a variable time, $\tau$, to allow inflow of $^{129}$Xe magnetization before acquiring 1D NMR profiles.

In addition to single spin-echo experiments, we used the CPMG technique [14] to measure the $^{129}$Xe spin coherence relaxation time ($T_2$) in the rock sample and diffuser plate, with $t_E = 2.1$ ms and one to sixteen 180° RF pulses prior to echo acquisition. For each point in the profile, we fit the amplitude decay, as a function of the number of echo loops before acquisition, to an exponential to yield $T_2(z)$. Fig. 2 shows example $T_2$ weighted profiles from such an acquisition sequence. After sixteen 180° pulses (33.6 ms), the xenon magnetization in the rock has completely dephased, but is clearly visible in the inflow tube and diffuser, thereby allowing us to define the start of the rock with an accuracy of about 1 mm (equivalent to the spatial resolution of the 1D NMR images).

To confirm the NMR experiments, each rock sample was also characterized independently using standard techniques by a commercial company [New England Research, White River Junction, VT]. Absolute porosity was determined using a gas pycnometer and calculated from gas pressure changes via Boyle's Law [15]. Permeability was measured via the steady-state gas flow method with a standard gas permeameter [2,3]. In these standard measurements, the accuracy of the absolute porosity value is generally accepted to be ~ 1%, and that of the permeability value ~ 10 – 20%.

### III. EFFECTIVE POROSITY MEASUREMENT AND RESULTS

To determine the effective porosity, we measured $^{129}$Xe NMR spin echo profiles from each sample under the condition of steady-state polarized gas flow. Fig. 3 shows examples of profiles acquired for the Fontainebleau sandstone and Austin Chalk samples. The amplitude of the profile at each point along the sample is proportional to the $^{129}$Xe gas density and spin polarization, the void space volume participating in gas flow, and the effect of spin coherence relaxation. As discussed in the next section, we corrected for gas density and polarization variation along the rock (see Fig. 3), such that the ratio of profile amplitudes in the diffuser plate (a medium of known porosity) and the rock was proportional to the ratio of the average void space volumes contributing to fluid flow in the two regions, with weighting factors accounting for $T_2$ relaxation. Our measurements of $T_2$ as a function of position along the sample yielded distinct values corresponding to $^{129}$Xe in the diffuser plate ($T_2^{dif}$ = 76.8 ms) and the rock cores ($T_2^{rock}$ = 4.59 ms in Fontainebleau sandstone and 12.3 ms in Austin Chalk). (These differences in $T_2$ arise from differences in magnetic susceptibility and hence background magnetic gradients, as well as differing wall interactions.)

Hence, we determined the effective porosity of the sample from the relation:

$$\phi_e = \frac{A_{dif}}{A_{rock}} \frac{S_{rock}}{S_{dif}} \frac{\exp(-t_E/T_2^{dif})}{\exp(-t_E/T_2^{rock})}, \qquad (3)$$

where $A_{dif}$ is the cross-sectional area of void space in the diffuser; $A_{rock}$ is the rock sample cross-sectional area; $S_{dif}$ and $S_{rock}$ are the NMR profile amplitudes in the diffuser and the rock respectively; and $t_E$ is the echo time used to acquire the profile. $^{129}$Xe spins located in isolated or dead-end pores larger than the one-dimensional gas diffusion length during $t_E$ (~ 50 $\mu$m for our experimental conditions) contributed no significant NMR signal to the effective porosity measurement. We determined the rock profile amplitude ($S_{rock}$) at a distance of 1 mm from the diffuser-rock interface on the upstream side of the gas flow. This 1 mm offset was chosen so that no $^{129}$Xe in the diffuser contributed to $S_{rock}$, and so that insignificant depolarization had occurred for the $^{129}$Xe in the well-connected pores that contribute to the rock's effective porosity and permeability. We also used the 1 mm offset point in the rock profile to calculate the correction for gas density and polarization variation along the rock. For additional 1 mm offsets, we found that the effective porosity derived from our measurements varied by less than other sources of uncertainty: < 3% fractional variation for the Fontainebleau samples, and < 6% fractional variation for the Austin chalk sample. Table 1 lists the effective porosities we determined for the Fontainebleau sandstone and Austin Chalk rock cores, as well as the absolute porosities determined using the gas pycnometer.

## IV. PERMEABILITY MEASUREMENT AND RESULTS

The permeability of a porous medium is generally determined from the volume flow rate of fluid under a given pressure gradient [3]. Here, we employed NMR imaging to determine the flow rate of laser-polarized $^{129}$Xe gas through reservoir rocks, using the pre-saturation method described above to image only xenon spins that flow into the rock during a defined period.

A one-dimensional NMR profile acquired with a spin-echo sequence provides a good representation of the spatial distribution of spin magnetization per unit length, $M(z)$, which can be expressed as [16]:

$$M(z) = A \cdot \phi_e \cdot n(z) \cdot \lambda \cdot p(z) \cdot (\gamma \hbar I), \qquad (4)$$

where $A$ is the cross-sectional area of the sample; $\phi_e$ is the effective porosity; $n(z)$ is the gas number density along the direction $z$ of gas flow; $\lambda$ is the isotopic abundance of $^{129}$Xe nuclear species in the Xe gas; $p(z)$ is the $^{129}$Xe spin polarization; and $\gamma \hbar I$ is the spin magnetization per polarized $^{129}$Xe atom (nuclear spin $I = 1/2$). On length scales > 1 mm, where $\phi_e$ is spatially uniform for the rocks we studied, only the number density, $n$, and polarization, $p$, are spatially dependent.

Assuming uniform laminar flow (a reasonable assumption for the very low Reynolds number, Re ~ $10^{-5}$, of these experiments, the 1D distribution of the gas number density inside the rock sample is given by [17]:

$$n(z) = \frac{1}{k_B T}\sqrt{P_i^2 - (P_i^2 - P_o^2)\frac{z}{L}}, \tag{5}$$

where $P_i$ and $P_o$ are the inlet and outlet gas pressures across the sample; $L$ is the sample length; $k_B$ is Boltzmann's constant; and $T$ is the gas temperature. The $^{129}$Xe spin polarization decreases along the flow path due to spin relaxation in the rock; the attenuation over a spatial displacement $dz$ is $\frac{dp(z)}{dz} = \frac{-1}{v(z)T_1(z)}$, where $v(z)$ is the spatially-dependent gas flow velocity and $T_1(z)$ is the spatially-dependent relaxation time. Applying Darcy's Law, for $v(z) = \frac{\vec{q}}{A\phi_e}$, one finds:

$$v(z) = -\frac{k}{\phi_e \mu}\frac{dP}{dz} = \frac{1}{2}\frac{k}{\phi_e \mu}\frac{P_i^2 - P_o^2}{LP(z)}. \tag{6}$$

where $P(z) = n(z)k_B T$, $\mu$ is the xenon viscosity and $k$ is the sample permeability Then the spatial dependence of polarization is found to be:

$$p(z) = p_0 \exp\left(-\int_0^z \frac{dz}{v(z)T_1(z)}\right) = p_0 \exp\left(-\frac{2\phi_e \mu}{k\beta}\frac{L}{P_i^2 - P_o^2}z\right), \tag{7}$$

where $p_0 = p(z=0)$ and $\beta$ relates the $^{129}$Xe $T_1$ to gas pressure via $T_1(z) = \beta P(z)$, assuming $T_1$ is dominated by wall relaxation and hence by gas diffusion to pore walls.

Combining Eqs. (4) to (7), the spatial dependence of the $^{129}$Xe spin magnetization per unit length in a porous rock can be written as

$$M(z) = A\phi_e \frac{1}{k_B T}(\gamma \hbar I)p_0 \sqrt{P_i^2 - (P_i^2 - P_o^2)\frac{z}{L}}\exp\left(-\frac{2\phi_e \mu}{k\beta}\frac{L}{P_i^2 - P_o^2}z\right). \tag{8}$$

Our permeability experiments were performed with a steady state gas flow, but with a zero initial $^{129}$Xe spin polarization, created by the application of a series of fast RF and magnetic field gradient pulses to spoil any spin polarization in the sample on timescales of 20 ms, which is much faster than the time for gas transport through the sample (on the order of 10 s). During a subsequent delay or propagation time, $\tau$, polarized gas entered the sample, after which we acquired an NMR profile. We repeated this process for different propagation times, thereby revealing the rate of flow of laser-polarized $^{129}$Xe gas through the rock's pore space. Example data for the Fontainebleau sandstone and Austin Chalk are shown in Fig. 4. The penetration depth, $\xi$, was calculated at each $\tau$ by dividing the total $^{129}$Xe NMR signal in the rock, determined from the MRI profile integrated over the rock length, by the signal amplitude at $z = 0$. We measured the inlet and outlet pressures $P_i$ and $P_o$ with bridge pressure sensors. (Typically, $P_i \approx 3.73$ bar and $P_o \approx 0.78$ bar for the low-permeability Austin Chalk, and $P_i \approx P_o \approx 4.33$ bar for the high permeability Fontainebleau.)  Also, we used the xenon viscosity $\mu = 2.324 \times 10^{-5}$ Kg/m-s, for the typical experimental temperature of 25°C [18]. (Gas viscosity is essentially independent of pressure and only minimally temperature dependent [18].)  Using Eqs. (5) and (6) as well as the ideal gas law ($P(z) = n(z) k_B T$), we derived a relation between the propagation time $\tau$ and penetration depth $\xi$:

$$\tau = \int_0^\xi \frac{dz}{v(z)} = \frac{4}{3} \frac{\phi_e \mu}{k} \frac{L^2}{(P_i^2 - P_o^2)^2} \left\{ P_i^3 - \left[ P_i^2 - (P_i^2 - P_o^2) \frac{\xi}{L} \right]^{\frac{3}{2}} \right\} . \tag{9}$$

From Eq. (9), we determined each rock's permeability, $k$, using the experimentally derived values for all other parameters.

For both the effective porosity and permeability measurements, we corrected for gas density and spin polarization variations as described in the following. We fit the $^{129}$Xe NMR profiles from each rock sample, measured with steady-state xenon flow and without prior polarization destruction (e.g., the profiles shown in Fig. 3), to Eq. (8). From these fits we determined the exponential decay rate, $\frac{2\phi_e \mu}{k\beta} \frac{L}{P_i^2 - P_o^2}$. Then, we removed the pressure and spin-relaxation dependence of each profile by dividing $S_{\text{rock}}$ (determined at a 1 mm offset from the diffuser-rock interface) by $\frac{1}{P_i} \sqrt{P_i^2 - (P_i^2 - P_o^2) \frac{z}{L}} \exp(-\frac{2\phi_e \mu}{k\beta} \frac{L}{P_i^2 - P_o^2} z)$, which is normalized to unity at 1 mm from the entrance of the rock sample. After this correction (see Figs. 3 and 4), profile amplitudes depended only on the fractional volume of the rock occupied by flowing laser-polarized xenon. Table 1 lists the permeabilities we determined for the two rock types, as well as the permeabilities obtained independently using the gas permeameter.

After determining the effective porosity and permeability of each sample, we determined the spin relaxation rate proportionality coefficient, $\beta$, from the exponential decay rate derived from fitting $^{129}$Xe NMR profiles to Eq. (8). For the Fontainebleau sample, the $^{129}$Xe spin relaxation time ($T_1$) could therefore be calculated from the nearly constant gas pressure in the sample ($P_i \approx P_o \approx 4.33$ bar), yielding $T_1 = 6.0 \pm 1.0$ s. To test the validity of Eq. (8) and the method of profile correction, we measured the $^{129}$Xe $T_1$ directly; we created a sealed container holding a large sample of Fontainebleau sandstone and thermally-polarized $^{129}$Xe at the same gas pressure used in the flowing, laser-polarized gas experiment. The result of $5.6 \pm 0.3$ s agrees well with the $T_1$ value derived from the flowing, laser-polarized gas experiment.

## V. DISCUSSION AND CONCLUSIONS

The laser-polarized noble gas NMR measurement of permeability agrees well with measurements made using the standard gas permeameter technique, for the representative high and low permeability rocks studied so far. For values above 500 mD and below 5 mD, the NMR method yields permeability results that are well within the uncertainty range of the values measured using the gas permeameter. In addition, the effective porosity simultaneously measured by the NMR technique shows that as permeability decreases, the effective porosity measured is a decreasing fraction of the absolute porosity determined by the standard gas pycnometer technique. Effective porosity (i.e., the volume fraction of pore spaces that contribute to fluid flow through the material) is always smaller than the absolute porosity of a sample, and is a parameter that is not easily, nor directly determined with other techniques. The Fontainebleau sandstone has a high permeability due to its well-connected pores with a narrow distribution of sizes, a fact that is consistent with the finding that the sample core has an effective porosity nearly as large as its absolute porosity. Conversely, Austin Chalk exhibits an effective porosity that is almost half the value of its absolute porosity, consistent with the knowledge of its very low permeability due to poor pore interconnectivity.

To perform NMR imaging of the penetrating inflow of laser-polarized xenon gas, the $^{129}$Xe spin decoherence time ($T_2$) in the rock samples must be sufficiently long (~ 2 - 5 ms), for a spin echo profile to be obtained without significant signal loss. For the Fontainebleau and Austin Chalk samples, both relatively free of paramagnetic impurities, this condition is easily satisfied, with the measured $^{129}$Xe $T_2$ at 4.7 T being 4.59 and 12.3 ms in the pores of the Fontainebleau and Austin Chalk, respectively. However, the paramagnetic impurities in many rocks, especially sandstones, will produce large magnetic field gradients when placed in magnetic fields $\geq 1$ T, these background gradients will significantly shorten the $^{129}$Xe $T_2$. One such example was Berea 100, a macroscopically homogeneous, high-permeability [4] sandstone with a narrow distribution of pore diameters of ~ 100 µm, but with a significant content of paramagnetic particles. In this sample, we were unable to measure a spin echo profile at 4.7 T from inflowing laser-polarized xenon, and hence could not measure the effective porosity or permeability. For such samples, it should be practical to operate at applied magnetic fields << 1 T for two reasons; (i) the magnetization of laser-polarized noble gas is determined by the optical pumping process external to the applied NMR magnetic field, $B_0$, i.e., the laser-polarization obtained is independent of $B_0$, whereas for thermally-polarized samples it is proportional to $B_0$; and (ii) the magnetic field gradients induced in porous media by magnetic susceptibility variations are greatly reduced for small $B_0$. We have previously demonstrated that NMR images of laser polarized noble gas can be acquired at applied field strengths as low as 20 G with resolution and SNR comparable to NMR images obtained at magnetic field strengths ~ 1 T [19,20]. Low-field NMR of flowing laser-polarized noble gas may allow effective porosity and permeability measurements in a wide array of porous samples using a simple, low-cost electromagnet. In addition to reservoir rocks, this technique may be applicable to ceramics, fluidized beds, filters and partially liquid-saturated porous media [21,22].

In conclusion, we have demonstrated the simultaneous measurement of permeability and effective porosity of high and low permeability oil-reservoir rock cores using NMR imaging of the penetrating flow of laser-polarized xenon gas. The method is accurate, with permeability results that cover a range of more than two orders of magnitude and agree well with the results from standard techniques. The method is also fast and reproducible: the procedure typically requires about 15 minutes, which is considerably less time-consuming than other NMR-based methods [17] and some standard techniques [6, 7]. The effective porosity measurements are consistent with expectations: the effective porosity is found to be nearly as large as the absolute porosity for the high permeability Fontainebleau sandstone, and to be significantly smaller than the absolute porosity for the lower permeability Austin Chalk sample. In future work, we will also investigate the utility of laser-polarized $^3$He gas to such measurements. In comparison to $^{129}$Xe, polarized $^3$He gas typically provides a larger NMR signal [19] and has weaker depolarizing wall interactions [23].

## ACKNOWLEDGEMENTS

We thank Dan Burns for technical assistance and liaison with New England Research. Additionally, we thank Ulrich Scheven and Schlumberger-Doll Research for the Teflon sample holder and rock samples. We acknowledge support by NSF grant CTS-9980194, NASA grant NAG9-1166 and the Smithsonian Institution.

# FIGURE CAPTIONS

Figure 1. Schematic diagram of the experimental apparatus. The 4.7 T magnet resides in a small RF shielded room. The remaining equipment was placed outside the room, beyond the 5 gauss line of the magnet. Narrow 1/8 inch ID Teflon tubing connected all pieces of the apparatus. The tubing length was approximately 2.5 m from the polarizer to the sample, and 5 m from the sample to the mass flow controller.

Figure 2. Example NMR profiles (i.e., 1D images) of laser-polarized xenon gas flowing through the Austin Chalk sample while applying a CPMG sequence before the image acquisition. The profiles include the regions occupied by the rock sample ($z \geq 13$ mm) and the Teflon diffuser plate (indicated by the dashed lines). The unattenuated profile ($t_E = 2.1$ ms) was obtained after only a single spin-echo before image acquisition. The attenuated profile ($t_E = 33.6$ ms) was acquired after sixteen 180° pulses, and is thus heavily $T_2$-weighted. In this profile, the $^{129}$Xe signal from the rock has completely dephased and is very small, while the $^{129}$Xe NMR signal from the diffuser plate remains significant. NMR profiles such as these allow $T_2(z)$ to be determined, and also permit unambiguous identification of the position of the rock core in the experimental apparatus.

Figure 3. Example NMR profiles (i.e., 1D images) of laser-polarized xenon gas flowing through samples of (a) Fontainebleau sandstone and (b) Austin Chalk, with both the gas flow rate and xenon magnetization in steady state. The profiles include the regions occupied by the rock sample ($z \geq 13$ mm) and the Teflon diffuser plate (indicated by dashed lines). The bold lines show the profiles corrected for gas density and polarization variation in the rock. For such typical NMR profiles, we averaged 32 signal acquisitions, each made with $t_E = 2.1$ ms, and achieved a 1D spatial resolution of approximately 1 mm.

Figure 4. Examples of $^{129}$Xe NMR profiles used in the permeability measurements: (a) Fontainebleau sandstone; (b) Austin Chalk. Profiles shown in solid lines correspond to different delay times, $\tau$, following a sequence of RF and gradient pulses to quench all xenon magnetization in the sample. The dash lines are profiles corrected for gas density and polarization variation.

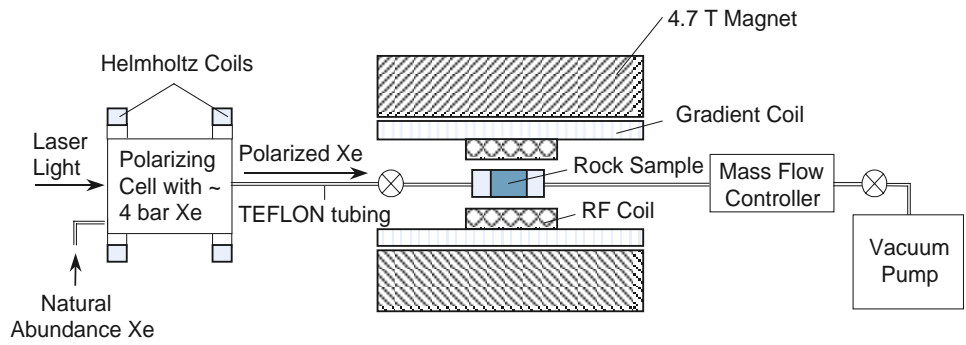

**Figure 1**

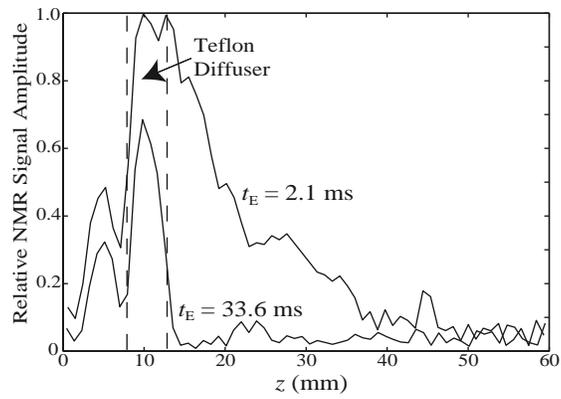

**Figure 2**

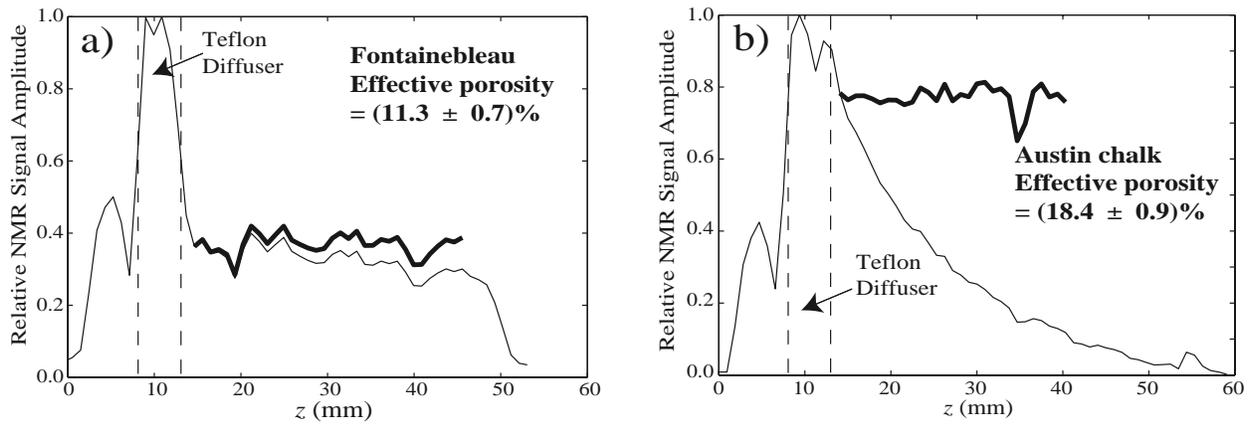

**Figure 3**

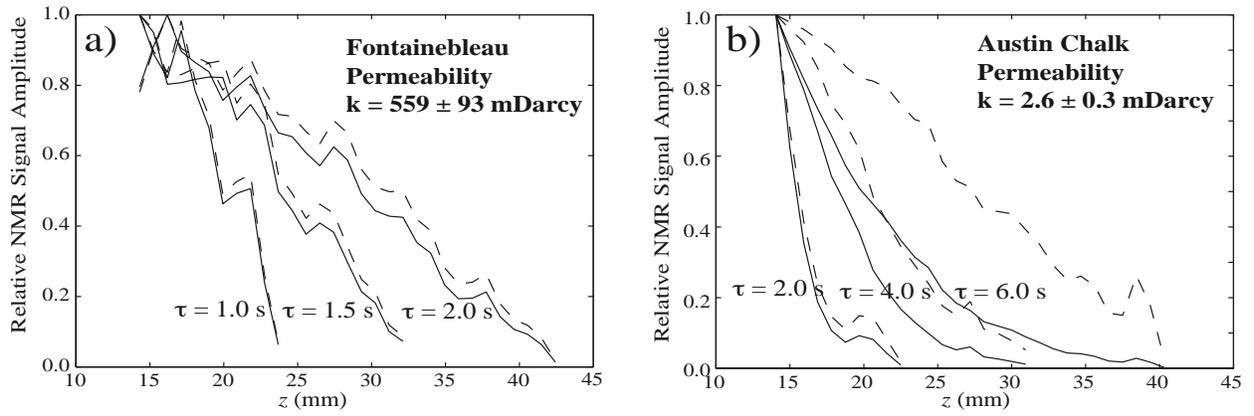

**Figure 4**

# Table 1

| Sample | Permeability (mD) | | Effective Porosity (%) | Absolute Porosity (%) |
|---|---|---|---|---|
| | LP-Xenon MRI | Gas Permeameter | LP-Xenon MRI | Gas Pycnometer |
| Fontainebleau | 559 ± 93 | 589 | 11.3 ± 0.7 | 12.5 |
| Austin Chalk | 2.6 ± 0.3 | 3.6 | 18.4 ± 0.9 | 29.7 |